\documentclass[aps,reprint,superscriptaddress,
nofootinbib,
amsmath,amssymb,
prc
]{revtex4-1}
\usepackage{natbib}
\usepackage{graphicx}
\usepackage{dcolumn}
\usepackage{bm}
\usepackage[T1]{fontenc}
\usepackage{ulem}
\usepackage{url}
\graphicspath{{./}}
\usepackage[usenames]{color}
\usepackage[colorlinks=true,allcolors=blue,dvipdfm]{hyperref}

\newcommand{\argmax}{\mathop{\rm arg~max}\limits}
\newcommand{\argmin}{\mathop{\rm arg~min}\limits}

\begin{document}

\preprint{...}

\title{Uncertainty quantification in nuclear shell model}


\author{Sota Yoshida}
\email{s.yoshida@nt.phys.s.u-tokyo.ac.jp}
\affiliation{Department of Physics, the University of Tokyo,~Hongo,~Bunkyo-ku,~Tokyo~113-0033,~Japan}

\author{Noritaka Shimizu}
\affiliation{Center for Nuclear Study, the University of Tokyo,~Hongo,~Bunkyo-ku,~Tokyo~113-0033,~Japan}

\author{Tomoaki Togashi}
\affiliation{Center for Nuclear Study, the University of Tokyo,~Hongo,~Bunkyo-ku,~Tokyo~113-0033,~Japan}

\author{Takaharu Otsuka}
\affiliation{RIKEN Nishina Center, 2-1 Hirosawa, Wako, Saitama 351-0198, Japan}
\affiliation{Department of Physics, the University of Tokyo,~Hongo,~Bunkyo-ku,~Tokyo~113-0033,~Japan}
\affiliation{Center for Nuclear Study, the University of Tokyo,~Hongo,~Bunkyo-ku,~Tokyo~113-0033,~Japan}
\affiliation{Instituut voor Kern- en Stralingsfysica, KU Leuven, B-3001 Leuven, Belgium}

\date{\today}

\begin{abstract}
The uncertainty quantifications of theoretical results are of great importance to make meaningful comparisons of those results with experimental data and to make predictions in experimentally unknown regions.
By quantifying uncertainties, one can make more solid statements about, e.g., origins of discrepancy in some quantities between theory and experiment.
We propose a novel method for uncertainty quantification for the effective interactions of nuclear shell-model calculations as an example.
The effective interaction is specified by a set of parameters, and its probability distribution in the multi-dimensional parameter space is considered.
This enables us to quantify the agreement with experimental data in a statistical manner and the resulting confidence intervals show unexpectedly large variations. 
Moreover, we point out that a large deviation of the confidence interval for the energy in shell-model calculations from the corresponding experimental data can be used as an indicator of some exotic property, e.g. $\alpha$ clustering, etc.
Other possible applications and impacts are also discussed.
\end{abstract}

\maketitle


{\it Introduction.} Many modern theoretical models have been developed by iterative cycles of solving "forward modeling problem" and "inverse modeling problem"~\cite{NatBayes}, and it has provided deep insight into phenomena of interest.
In particular, such cycles have been playing key roles in nuclear physics, because the fundamental interaction, the nuclear force,
is still uncertain due to its complexity including the non-perturbative character in the low-energy regime.
In applications of such theoretical models, one has to solve the inverse modeling problem, i.e. the optimization
problem to minimize deviations of model estimates from observations.
However, if one uses a model with so-called point estimation of the optimal input parameters, there is a risk of overfitting to the given observations and of lacking the generalization ability to the data not taken into the optimization process.
Therefore, it is desired that any theoretical model should be properly accompanied with uncertainty estimates, see e.g. Editorial on Physical Review~A~\cite{EditorialPRA}.
Theoretical studies in nuclear theory with uncertainty estimates are being expanded, e.g., mean-field calculations \cite{UQ_Nskin,UQ_RMF,UQ_NskinEDF,UQ_SkyrmeEDF,Dobaczewski2014,UQ_HFB,UQ_CEDF,UQ_McDonnell}, nuclear potentials~\cite{UQpot_1,*UQpot_1e, UQpot_2,UQ_chiEFT,UQ_PRX}, etc.

The shell-model calculation, which is called the configuration interaction method in other fields of science, has been providing successful and systematic descriptions of a wide variety of properties of light to medium-mass nuclei.
Therefore one can expect that this model well approximates wave functions of nuclei, see e.g. Refs.~\cite{Brown_Rev,RMP_Caurier,RMP_OTsuka}.
Conventional shell-model Hamiltonians have been constructed by the $G$-matrix theory~\cite{GKB,GMorten} with minor phenomenological corrections.
Not only from phenomenological perspectives, but also from microscopic viewpoints, shell model is playing a key role with recent developments in nuclear forces from chiral effective field theory (chiral EFT)~\cite{EMrev,EGMrev} and many-body methods to derive shell-model effective interactions for a physically motivated model space~\cite{VS-IMSRG1,VS-IMSRG2,VS-IMSRG3,CCEI1,CCEI2,CCSM}.
The latter approaches combining {\it ab initio} methods and shell-model calculations can act as a foothold for better understanding of the nuclear potential.
Under these circumstances, it is an urgent task to assess the validity of nuclear shell model through the evaluations of their uncertainties stemming from input effective interactions.
This is because by performing detailed analyses of uncertainties coming from input parameters in the theory one can make more solid statements, such as on origins of the discrepancies between theoretical predictions and experimental observations, or on which observables should be reproduced by the theory within a given model space.
In this study, we provide, for the first time, uncertainty quantifications of shell-model spectra in a statistical manner.


{\it Inference on effective interactions.} In what follows, we introduce a way to quantify uncertainties in shell-model results by introducing Bayesian inference on input parameters and then
we consider the shell-model results with ensembles of effective interactions.
It provides rich information about the validity of theoretical estimates, while in conventional shell-model studies results with only one or a few parameter sets are compared with experimental values.

We take the $0p$-shell space on top of the ${}^{4}$He core as the model space.
The $0p$ shell consists of two orbitals $0p_{1/2}$ and $0p_{3/2}$ and these are abbreviated as $p_{1/2}$ and $p_{3/2}$ in this paper.
In this case, the shell-model Hamiltonian contains the 17 parameters with the isospin symmetry: 2 single-particle energies (SPEs) and 15 two-body matrix elements (TBMEs).
We do inference on these parameters under a given set of experimental energy values for some $p$-shell nuclei
and regard the parameters not as points but as probability distributions over 17-dimensional parameter space.
All calculations within the $0p$-shell space can be done by an exact diagonalization method with relatively low computational costs.
In this case, since there is no uncertainty coming from the many-body calculation, i.e. the diagonalization process, we can extract a theoretical uncertainty coming merely from input parameters.

In Bayesian analysis the parameter distribution is described by the posterior distribution,
i.e. the conditional probability distribution under observation of data.
This posterior is obtained through Bayes' theorem,
\begin{align}
P(\boldsymbol{\theta} | D ) = \frac{P(D|\boldsymbol{\theta}) P(\boldsymbol{\theta})}{P(D)} \propto P(D|\boldsymbol{\theta}) P(\boldsymbol{\theta}),
\label{eq:Bayes}
\end{align}
where $\boldsymbol{\theta}$ is a set of multidimensional parameters.
In this study, $\boldsymbol{\theta}$ corresponds to the 17-dimensional parameters for the $p$-shell effective interactions.
The $D$ in the equation above denotes the data set taken into account for the parameter inference.
We take the uniform distribution as the prior, $P(\boldsymbol{\theta}) \propto 1$, and the ordinary likelihood function,
\begin{align}
P(D|\boldsymbol{\theta}) = \exp{( -\chi^2 (\boldsymbol{\theta})/2)},
\label{eq:likelihood}
\end{align}
with the squared errors,
\begin{align}
\chi^2(\boldsymbol{\theta}) \equiv  \sum^{N_D}_{n=1} 
\left( \frac{ \mathcal{O}^{\mathrm{exp}}_n - \mathcal{O}_n^\mathrm{th} [\boldsymbol{\theta}]}
{\Delta \mathcal{O}} \right)^2.
\label{eq:error}
\end{align}
Here $N_D$ denotes the number of data,
$\mathcal{O}^{\mathrm{exp}}_n$ is an experimental value for the observable labeled by $n$,
and $\mathcal{O}_n^\mathrm{th} [\boldsymbol{\theta}]$ is the corresponding theoretical results with $\boldsymbol{\theta}$.
We note that since we now assume the isospin symmetry in effective interactions,
the Coulomb corrections to energy values for $\{\mathcal{O}_n^\mathrm{th} [\boldsymbol{\theta}]\}$ are calculated in accordance with Ref.~\cite{CK}.
The $\Delta \mathcal{O}$ in the Eq.~\eqref{eq:error} is the typical error of the observables and this should contain both theoretical and experimental errors in general.
However, the experimental error is negligible in the present work.
For $\{ \mathcal{O}^{\mathrm{exp}}_n \}$, we use a fixed 33 data for the energy values of ground and excited states of the $p$-shell nuclei throughout this work.
Those are essentially the same set as used in the work by Cohen and Kurath~\cite{CK} with the exception of a few updated data.
The interest reader is referred to the Supplemental Material~\cite{Supple_SY}.
If one intends to change the data set in the fitting procedure, one can modify the parameter distributions relatively easily by regarding the previously obtained posterior as the prior and by doing the similar inference for the data added.
Furthermore, if one makes pseudo data for unknown states, one can estimate their impacts on parameters in the same manner.
This flexibility is a benefit of introducing Bayesian inference.

We typically have two classes to evaluate the posterior $P(\boldsymbol{\theta} | D)$. One is the asymptotically exact method, Markov Chain Monte Carlo (MCMC)~\cite{MCMChandbook}.
If one could achieve an infinite number of iterations for the MCMC, one would obtain the samples, i.e. the ensemble of effective interactions that are exactly obeying posterior distributions.
However, in the case of highly multimodal or too steep true posterior, it is hard to obtain fully converged results by the MCMC with currently available computer resources.
The other class is the approximation schemes, which are literally approximation methods for posteriors; one can write down a posterior in a simple and closed form like a Gaussian distribution.
In this class, computational costs are in general much less than those for the MCMC.
The advantages and disadvantages for these two classes are complementary.
In this work, we employ the Laplace approximation (LA), which belongs to the latter class.
In the problems of interest, the LA tends to search a wider region in the parameter space than the MCMC,
because the $\chi^2$ potential is considerably steep around the global minimum.
This means that uncertainties in the parameters evaluated by LA samples are larger than those obtained by the MCMC.
We adopt the LA because it mitigates the risk of too definitely estimating the uncertainty.

When applying the LA, the posterior is approximated by multivariate Gaussian around the so-called Maximum A Posteriori (MAP) estimate:
\begin{align}
\boldsymbol{\theta}_{\mathrm{MAP}} & \equiv \argmax_{\boldsymbol{\theta}} P(D|\boldsymbol{\theta})P(\boldsymbol{\theta}) =  \argmin_{\boldsymbol{\theta}} \chi^2(\boldsymbol{\theta}), \label{eq:MAP}\\
P(\boldsymbol{\theta}|D) &  \approx 
\frac{|\boldsymbol{A}|}{\sqrt{(2\pi)^k }}
\exp{
\left(-\frac{1}{2} \bar{\boldsymbol{\theta}}^T \boldsymbol{A}
\bar{\boldsymbol{\theta}}
\right)},
\label{eq:posterior}
\end{align}
with $\bar{\boldsymbol{\theta}}=\boldsymbol{\theta}-\boldsymbol{\theta}_{\mathrm{MAP}}$ and the Hessian matrix $\boldsymbol{A}$
\begin{align}
\boldsymbol{A} &= - \nabla \nabla \log P(\boldsymbol{\theta}|D)|_{\boldsymbol{\theta}=\boldsymbol{\theta}_{\mathrm{MAP}}}.
\label{eq:Hessian}
\end{align}
In Eq.~\eqref{eq:MAP}, we used Eqs.~(\ref{eq:Bayes}-\ref{eq:likelihood}) and  $P(\boldsymbol{\theta})\propto 1$.
Each element of the Hessian matrix $\boldsymbol{A}$ is written as
\begin{align}
A_{ij}& = 
\sum^{N_D}_{n=1} 
\frac{1}{(\Delta \mathcal{O})^2}
\frac{\partial \mathcal{O}^\mathrm{th}_n[\boldsymbol{\theta}] }{\partial \theta_i}
\frac{\partial \mathcal{O}^\mathrm{th}_n[\boldsymbol{\theta}] }{\partial \theta_j} \nonumber \\  
& \hspace{2cm} +
\sum^{N_D}_{n=1}  
\frac{( \mathcal{O}^\mathrm{th}_n[\boldsymbol{\theta}]  - \mathcal{O}^{\mathrm{exp}}_n)}{(\Delta \mathcal{O} )^2} 
\frac{\partial^2  \mathcal{O}^\mathrm{th}_n[\boldsymbol{\theta}] }{\partial \theta_i \partial \theta_j}.
\label{eq:1stder}
\end{align}

The first derivative of the term is obtained by the Hellman-Feynmann theorem~\cite{HF}
and the second derivative term above is numerically evaluated using the finite differences of the first derivative terms,
\begin{align}
\frac{\partial^2  \mathcal{O}^\mathrm{th}_n[\boldsymbol{\theta}]}{\partial \theta_i \partial \theta_j} = \frac{1}{2\epsilon}
\left\{
\frac{\partial \mathcal{O}^\mathrm{th}_n[\theta_j^+ ] }{\partial \theta_i}
-
\frac{\partial  \mathcal{O}^\mathrm{th}_n[\theta_j^- ] }{\partial \theta_i}
\right\}
+ \mathcal{O}(\epsilon^2),
\label{eq:2ndder}
\end{align}
where $ \mathcal{O}_n[\theta_j^ {\pm}]$ denotes the $\mathcal{O}_n$ value evaluated by the parameters whose $j$-th components are slightly shifted by a small value $\pm \epsilon$ from the $\boldsymbol{\theta}_{\mathrm{MAP}}$.

While we use the uniform prior and the LA for the posterior, (this particular choice reduces the parameter inference to the problem of chi-square minimization and covariance matrix analysis as in Refs.~\cite{
UQ_Nskin,UQ_RMF,UQ_NskinEDF,Dobaczewski2014,UQ_HFB,UQ_CEDF,UQ_McDonnell,UQpot_1,*UQpot_1e, UQpot_2,UQ_chiEFT,UQ_PRX}), the previously mentioned benefit of Bayesian inference could be achieved in future works by a fully Bayesian treatment of posterior distributions.

\begin{table}[b]
\caption{The root mean square (RMS) errors of energies for the 33 data in fit with the our optimized interaction and Cohen-Kurath interactions.  All errors are in the unit of MeV. \label{tab:chisq}}
\begin{ruledtabular}
\begin{tabular}{lccccc}
   &    & $\boldsymbol{\theta}_{\mathrm{MAP}}$& CKpot & CKtb1 & CKtb2 \\ \hline
 total RMS & & 0.35 & 0.57 & 0.47 & 0.54 
\end{tabular}
\end{ruledtabular}
\end{table}

The concrete procedures to quantify the uncertainties in shell-model calculations by the LA are the following:
(i) to search the optimal interaction, i.e. the $\boldsymbol{\theta}_{\mathrm{MAP}}$ in Eq.~\eqref{eq:MAP}, with respect to the given data set by means of optimization methods,
(ii) to calculate the Hessian matrix according to Eqs.\eqref{eq:Hessian}-\eqref{eq:2ndder},
and (iii) to generate a sufficiently large number of the parameter set $\{ \boldsymbol{\theta} \}$ obeying the posterior defined in Eq.~\eqref{eq:posterior} and to evaluate statistical quantities for observables of interest.
For (i), we prefix the optimal interaction by Stochastic Gradient Descent, see e.g.~\cite{SGD_RM}.
The root mean square errors for the given data set are summarized in Tab.~\ref{tab:chisq} in comparison with the results for three sets of the interaction proposed by Cohen and Kurath (CK)~\cite{CK}.
We note that the mass-dependence of the form $(A/6)^{-0.3}$ is introduced for TBMEs while three CK interactions are mass-independent.
Three CK interactions, denoted CKpot, CKtb1, and CKtb2, correspond, respectively, to (8-16)POT, (8-16)2BME, and (6-16)2BME, in the original work. Here, the numbers in the parentheses denote the range of the mass number used in fit.
In what follows, we use the total root mean square error for the optimal interaction as the typical error in Eq.~\eqref{eq:error}, $\Delta \mathcal{O} = 0.35$ MeV.
In the procedure (iii), we perform shell-model calculations with 50,000 LA samples.
This number is large enough to suppress the error coming from stochastic choices of samples;
The typical error in the mean values of energy eigenvalues is less than $0.1\%$.

\begin{figure}
\centering{
\includegraphics[width=8.6cm]{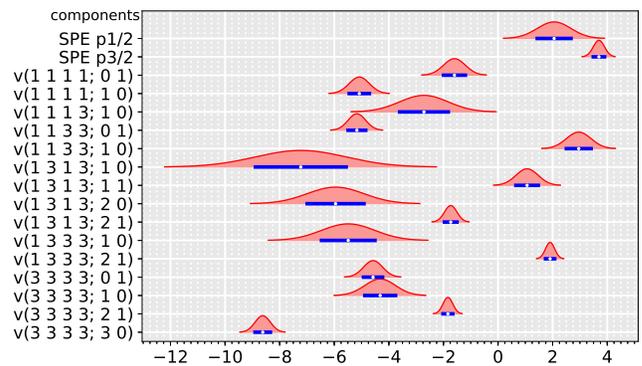}
\caption{
The marginal distributions of 17 parameters for the $p$-shell, in MeV. 
Red curves and blue bars denote the $3\sigma$ and $1\sigma$ confidence intervals, respectively.
Each plot is scaled to the same height for aesthetic purposes.
White dots denote the medium of the distribution and, namely, the $\boldsymbol{\theta}_\mathrm{MAP}$.
The TBMEs are abbreviated as v$(a b c d; JT)$, and the orbits are labeled by $1=p_{1/2}$ and $3=p_{3/2}$. \label{fig:comp}}
}
\end{figure}

{\it Results.} We show the marginal distributions of the 17 parameters for the $p$-shell interaction in Fig.~\ref{fig:comp}.
Red curves with the shaded area and blue bars respectively denote $3 \sigma$ and $1 \sigma$ confidence intervals for the parameter distributions.
The $1\sigma$ deviations of parameters are given by the square root of diagonal components of $\boldsymbol{A}^{-1}$.
The white dots plotted at the medians of Gaussian distributions correspond to the parameter values of the MAP estimation.
Some TBMEs, whose total angular momentum $J$ and total isospin $T$ are $(J,T)=(1, 0)$, show relatively large uncertainties.
If we take Gamow-Teller transition strengths or electromagnetic observables into fit, which are sensitive to those parameters, we expect that the uncertainties would become smaller.
However, this brings additional issues on the quenching factor or the effective charges.
In this work, we do not enter into the detail of those quantities for the sake of simplicity.

\begin{figure*}
\centering{
\includegraphics[bb=  0 0 721 219,width=17.2cm]{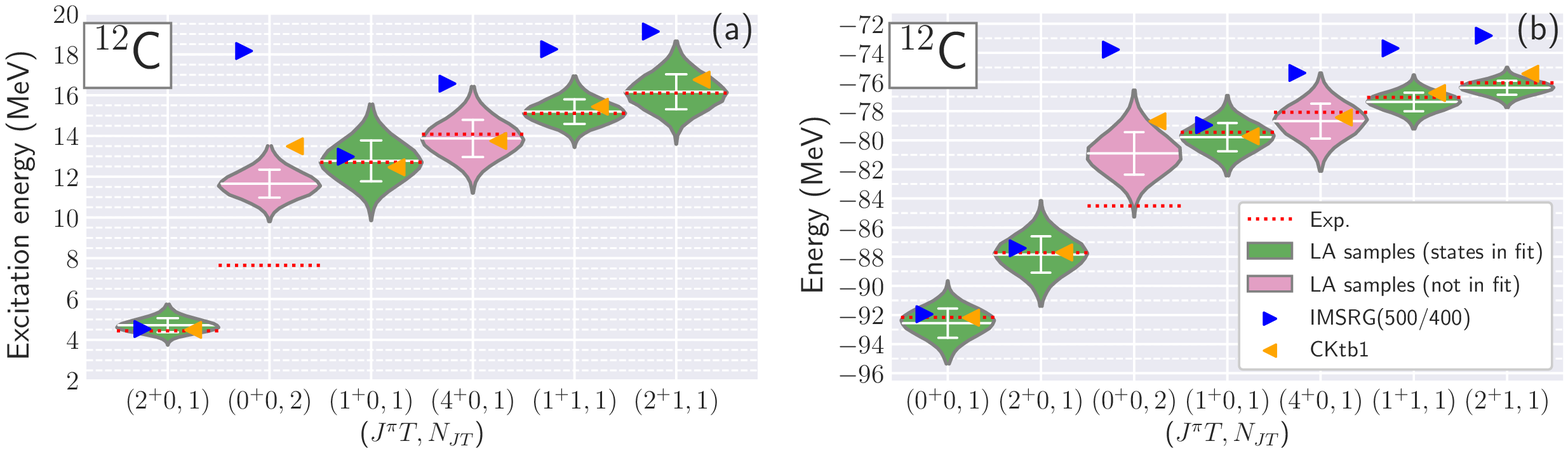}
\includegraphics[bb=  0 0 721 219,width=17.2cm]{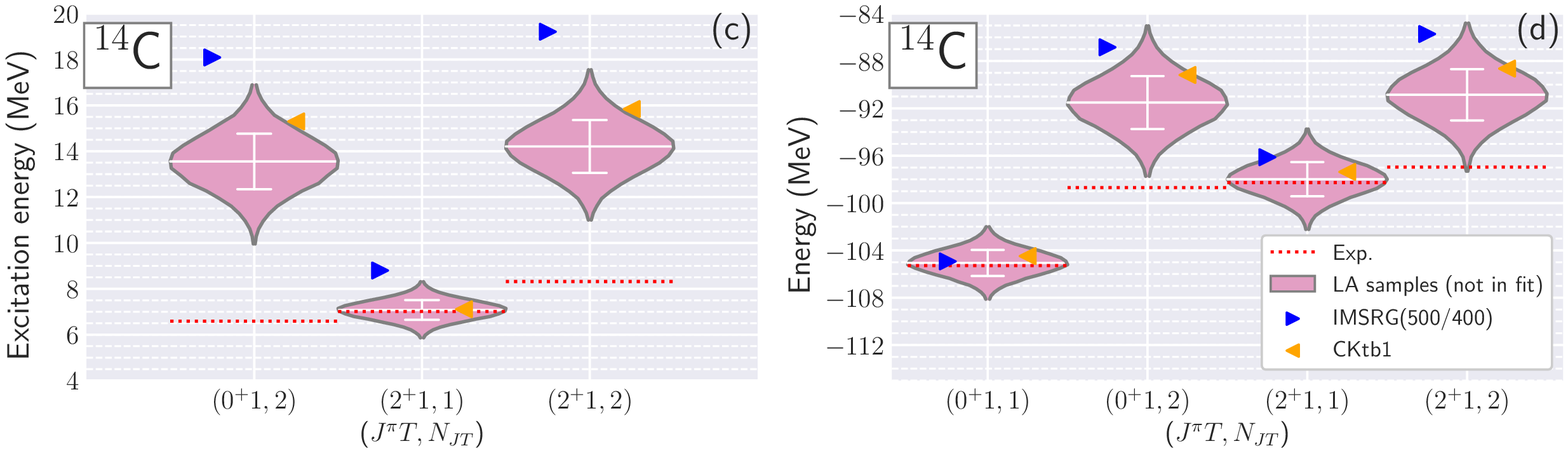}
\caption{
Excitation energies (left column) and energies (right column) of ${}^{12}$C (panels (a) and (b)) and ${}^{14}$C ((c) and (d)).
Each state is classified by its $(J^\pi T, N_{JT})$, where $(J^\pi T, N_{JT})$ stands for the $N$-th lowest states with the total angular momentum $J^\pi$ and the total isospin $T$.
Experimental data on ENSDF~\cite{ENSDF} are shown by horizontal dotted lines in red.
Corresponding theoretical results with the LA samples are summarized in violin plots in green or pink.
Here green and pink ones corresponds to the states in fit and not in fit, respectively.
In each violin plot, mean value and $1\bar{\sigma}$ error are shown by horizontal line and error bar in white, respectively, where $\bar{\sigma}$ means standard deviations of all results.
Results by a CK interaction and the VS-IMSRG are also shown.
\label{fig:spectra}}}
\end{figure*}

In Fig.~\ref{fig:spectra}, we show some selected results of energy spectra.
Theoretical results by the LA samples are shown by so-called violin plots in comparison with (i) experimental data~\cite{ENSDF}, (ii) the results by one of the CK interactions~\cite{CK}, and (iii) those by the microscopically derived interaction by the valence-space in-medium similarity renormalization group (VS-IMSRG)~\cite{VS-IMSRG3}.
The height and width of violins show, respectively, 3$\bar{\sigma}$ confidence intervals and appearance frequencies of the quantities with respect to all the 50,000 LA samples.
Here we use the notation $\bar{\sigma}$ for standard deviations of the results with all the LA samples in order to distinguish from the ordinary statistical term $\sigma$ for Gaussian distributions.
Their mean values (horizontal solid lines) and 1$\bar{\sigma}$ confidence intervals (error bars) are shown in violin plots.
Experimental data and theoretical results are classified with the angular momentum $J$, the parity $\pi$, and the isospin $T$.
We note that the total isospin is not determined in some data, in which cases most plausible values are taken from the corresponding theoretical results.
The CKtb1 was determined from the almost same data as the present study, and gives the minimum total root mean square errors among the three CK interactions (see Tab.~\ref{tab:chisq}).
Excitation spectra like Fig.~\ref{fig:spectra} (a, c) are convenient for comparisons with experimental studies with gamma-ray measurements, e.g. to determine which states are relevant to newly measured gamma rays.
One can also work with energy eigenvalues, as shown in Fig.~\ref{fig:spectra} (b, d).
These are convenient for purposes to see systematic deviations (e.g. with respect to the mass numbers) of shell-model results from experimental data. 


Hereafter, we consider 82 low-lying states of the $p$-shell nuclei in order to confirm that the parameter distributions are not overfitted to the 33 states. We mainly focus on the energy spectra for ${}^{12,14}$C in Fig.~\ref{fig:spectra} as a representative example.
By taking a close look at these results, we can make some important remarks.
Firstly, we can see significant variances in the height of confidence intervals.
This enables us to quantify the relative reliabilities of the theoretical estimates.
The overall scale of the confidence intervals is partly controlled by the typical error in the Eq.~\eqref{eq:error}, but the relative ratio of the height may suggest complicated correlations involving parameters and/or many-body structure.
Secondly, most of the states are described with the LA samples within $1\bar{\sigma}$ deviations, as shown in Fig.~\ref{fig:spectra}.
Indeed, most of the energy values for the 82 states are reproduced by the LA samples within $1\bar{\sigma}$ error, while there is a small number of exceptions as we discuss below.
This fact demonstrates the generalization ability of shell-model calculations to the data not taken into the parameter inference.
Thirdly, one can find a few states in Fig.~\ref{fig:spectra} whose mean energy values differ from the corresponding experimental data by more than $2\bar{\sigma}$:
the second $0^+$ state of ${}^{12}$C (Hoyle state) and the second $0^+$ and $2^+$ states of ${}^{14}$C.
Such states can be found in ${}^{14}$C(2), ${}^{12}$C(1), ${}^{10}$B(1), ${}^{12}$Be(1), ${}^{10}$Be(1), ${}^{7}$Li(1), and ${}^{6}$Li(1) of the 82 states, where the numbers in parentheses denote the number of states showing the deviation from experimental data by more than $2\bar{\sigma}$.
For the states listed above, it has been already suggested that the $p$-shell space is insufficient to describe these states, see e.g.~\cite{psdWB,Chernykh07,Enyo_1,*Enyo_2,Maris09,Epelbaum11,Carlson15,Fortune78,Wolters90,Nav00}.
We propose that these large discrepancies between the confidence intervals and experimental data can be interpreted as an indicator of exotic structures, e.g. $\alpha$ clustering, intruder configurations, core excitations, etc.
Although we cannot give a criteria for exotic structures, we can deduce which states are likely to have exotic structures by looking at the relative size of the discrepancies between confidence intervals and observations.
This is an important outcome of the evaluation of uncertainties in shell-model calculations.


Even if there is no preliminary knowledge of which states have exotic structures,
one can quantify how the states are to be (or not to be) taken into account through fitting procedures.
For example, if one adds the Hoyle state into the data set in fit, the total root mean square error becomes worse by $\sim 39\%$ for the 74 states which are reproduced within $2\bar{\sigma}$ error by the already obtained LA samples with the 33 data.
In this way, one can identify certain states which should not be included in the fitting procedure, and then update the optimal interaction to more utilitarian one.
Furthermore, we can assess the validity of the LA by looking into the shapes of violin plots.
If the LA failed to capture the global structure of $\chi^2$, i.e. in the case that there are many local minima or plateaus in the $\chi^2$-potential over the parameter space, it is expected that the mean values of observables calculated with the LA samples will deviate from the evaluation with the MAP estimate and that the shape of violin plots will become asymmetric.
However, the deviation is not significant and the shapes of violin are almost symmetric: Regarding all the 82 states, the typical discrepancy in energy values between a result by the $\boldsymbol{\theta}_\mathrm{MAP}$ and a mean value of the results with the LA samples is about $0.1\%$ (see~\cite{Supple_SY}).



Finally, the methodology discussed in this study would be useful for microscopic understandings too.
In the current microscopic calculations like the VS-IMSRG and the coupled-cluster method, the input nuclear potential is regarded as the dominant source of uncertainty~\cite{HH_rev1}.
The capability of shell-model calculations is important
to know which states are likely to be described by future microscopically derived effective interactions.
By visualizing the confidence intervals of shell-model calculations, there is also a possibility to figure out some missing contribution or higher order terms which are not to be omitted in modern nuclear forces and many-body methods.
For example, if the considered states have large deformations and are located in the confidence intervals of the shell-model calculation with phenomenologically constructed effective interactions, one may need to take account of higher order couplings between a valence space and an outer space.

{\it Conclusions.} We have proposed a novel method to quantify uncertainties in shell-model calculations.
By introducing violin plots, we have visualized the unexpectedly large variations of confidence intervals which differ by states. These plots can be used to quantify the agreement with experimental data in a statistical manner.
We have also proposed that a large deviation of theoretical confidence intervals from the corresponding experimental data can be regarded as an indicator of some exotic structure like clustering, intruder configurations, core excitations, etc.

The methodology discussed in this work may have various possible future applications.
By implementing Bayesian inference on the parameters, we can sequentially construct shell-model effective interactions while sorting out data.
This would be useful especially for heavier regions where even one iteration for the optimization process is computationally demanding.
It would be also interesting to extend this analysis to the valence-space Hamiltonians based on chiral EFT potentials for a valence space~\cite{Huth}.
One can also extend this study to include electromagnetic observables or Gamow-Teller transition strengths into the data set.
That is one of the future perspectives of this work and of great importance for astrophysical applications.

If these uncertainty estimates could be done for any other phenomenological models in which theoretical uncertainties have not been discussed in detail, it may provide new insight into a wide variety of phenomena in nuclear physics.

{\it Acknowledgments.} We thank Achim Schwenk and Yutaka Utsuno for fruitful discussions and comments.
We also thank Patrick Copinger for carefully reading the manuscript.
This work was supported in part by JSPS KAKENHI (Grant Numbers 17J06775 and 17K05433),
JSPS Overseas Challenge Program for Young Researchers (No. 201880031),
the priority issue 9 to be tackled by using Post-K computer,
and the CNS-RIKEN joint project for large-scale nuclear structure 
calculations. 

\normalem

\bibliography{pshl_main}







\end{document}